
\input jnl.tex
\refstylePR
\input reforder.tex
\vglue 1. truein
\title
{
On the Stability of Electroweak Strings
}
\author
{Margaret James}
\affil
{
D.A.M.T.P., Silver Street,
University of Cambridge, Cambridge, CB39EW, U.K.
}
\smallskip
\author
{Leandros ${\rm Perivolaropoulos^*}$}
\affil
{
Division of Theoretical Astrophysics,
Harvard-Smithsonian Center for Astrophysics,
60 Garden Street, Cambridge, MA 02138.
}
\smallskip
\author
{Tanmay Vachaspati}
\affil
{
Tufts Institute of Cosmology, Department of Physics and Astronomy,
Tufts University, Medford, MA 02155.
}

\abstract
\doublespace
We map the parameter space that leads to stable Z-vortices in the
electroweak model. For $sin^2
\theta_W = 0.23$, we find that the strings are unstable for a
Higgs mass larger than 24 GeV. Given the latest constraints
on the Higgs mass from LEP, this shows that, if the standard
electroweak model is realized in Nature, the Z-vortex
(in the bare model) is unstable.

\endtopmatter

The presence of vortices in a physically realized particle physics
model is an exciting prospect.
The presence of such vortices in the
electroweak model is even more exciting since it immediately opens
up the possibility of new signatures in accelerator experiments
at accessible energies.
If these signatures are found, it would be the
first time a coherent state would have been detected in particle
physics.

In previous papers\refto{tv91, tv92}
(see also Refs.\cite{com1, tvaa, mh}), one of us had
shown the existence of vortex
(/string) solutions in the Weinberg-Salam model\refto{swas} and had also
discussed the stability of the solutions.
An existence proof of the stability was provided and
it was clear that the stability crucially depends
on the values of the
parameters in the model. In this paper\refto{mistake}, we
map the range of parameters for which the electroweak string
solution is stable and, in particular, we study the physical case
in which $sin^2 \theta_W = 0.23$, $m_Z = 92 GeV$ and $m_H > 57 GeV$
(where $m_H$ is the mass of the Higgs particle).

Our approach to the problem is to consider the variation in the
energy up to second order in the perturbations of the fields about the
vortex solution. In
principle, there are four scalar fields and four vector fields
each with four components. This makes a total of twenty fields, each
of which has to be perturbed. However, remarkably, we are able to reduce the
problem down to only {\it one} field. For this
perturbation mode, we construct a Schrodinger
equation and numerically find the range of parameters for which there is
no bound state. This gives us the
parameter values for which the vortex solution is stable to small
perturbations.

We will use the standard notation
defined in Ref.\cite{jct}. In addition, we make the usual definitions:
$
Z^\mu \equiv cos\theta_W W^{\mu 3} - sin\theta_W B^\mu,
$
$
A^\mu \equiv sin\theta_W W^{\mu 3} + cos\theta_W B^\mu
$
where,
$
tan\theta_W \equiv g'/g.
$
Also,
$
\alpha \equiv \sqrt{ g^2 + {g'} ^2 }
$
where $g$ and $g'$ are the couplings of the $W_\mu^a$ and $B_\mu$ gauge
fields to the Higgs field.

The energy functional for static field configurations in
the Weinberg-Salam model is:
$$
E = \int d^2 x dz \left [
          \fourth G_{ij} ^a G_{ij} ^a + \fourth F_{Bij} F_{Bij}
          + (D_j \phi ) ^{\dag} (D_j \phi ) +
            \lambda (\phi ^{\dag} \phi - \eta ^2 /2 )^2
              \right ]
\eqno (1)
$$
where, $i,j,a = 1,2,3$. The integral over
the $z-$coordinate has been shown
explicitly since we shall show that
it is sufficient to consider field configurations in the $xy-$plane alone.

The vortex solution \refto{tv91, tv92} that extremizes the above energy
functional is:
$$
\eqalign {
W^{\mu 1} = 0 & = W^{\mu 2} = A^\mu , \ \ \
Z^\mu = [ A^\mu ] _{NO} = - {{v_{NO} (r)} \over r} {\hat e}_\theta
\cr
&
\phi = f_{NO} (r) e^{im \theta } \Phi \ , \ \ \
\Phi \equiv \pmatrix{0\cr 1\cr}
\cr
}
\eqno (2)
$$
where, the coordinates $r$ and $\theta$ are polar coordinates
in the $xy-$plane. The integer $m$ is the winding number of the
vortex and, here, we shall restrict ourselves to the case
$m=1$. The
subscript $NO$ on the functions $f$ and
$A^\mu$ means
that they are identical to the corresponding functions found
by Nielsen and Olesen\refto{hnpo} for the usual Abelian-Higgs string.
(We now drop the subscript $NO$ on the functions $f$ and
$v$ for ease of writing.)
These functions are given by the equations of motion:
$$
f'' + {{f'} \over r} - \left ( 1- {e \over 2} v \right ) ^2
                             {f \over {r^2}}
   - 2 \lambda \left ( f^2 - {{\eta ^2} \over 2} \right ) f = 0
\eqno (3)
$$
$$
v'' - {{v'} \over r} + e \left ( 1 - {e \over 2} v \right ) f^2 = 0
\eqno (4)
$$
where primes denote differentiation with respect to $r$. The functions
$f$ and $v$ also satisfy the boundary conditions:
$
f(0) = 0 = v(0), \ \ \  f(\infty ) = {\eta \over {\sqrt{2}}} , \ \ \
v(\infty ) = {2 \over \alpha}.
$
The string solutions resulting from these equations have been studied
previously by several authors in
a lot of detail. A sample of these papers may be found in the
collection of Ref. \cite{gscr}.

We shall now study the stability of the vortex solution given in eq. (2)
by considering infinitesimal perturbations around it and finding if the
variation in the energy is positive or negative. The perturbations are
time independent and therefore we can also set the zero components of
the gauge field to be zero.

Let us write
$$
\phi = \pmatrix{\phi_1\cr \phi_{NO} + \phi_2\cr}
\eqno (5)
$$
$$
Z^\mu = Z_{NO} ^\mu + \delta Z^\mu
\eqno (6)
$$
$$
T^1 \equiv diag( - cos2\theta_W , 1 ),
\eqno (7)
$$
and,
$$
{\bf d} _j \equiv ( \partial _j {\bf 1} + i\half \alpha T^1 Z_j ) \  .
\eqno (8)
$$
where, ${\bf 1}$ is the $2\times 2$ unit matrix.
Now, since we are considering perturbations on top of the vortex solution,
the fields $\phi_1$, $\phi_2$, $\delta Z^\mu$, ${W}^{\mu  {\bar a}}$
($\bar a = 1,2$)
and $A^{\mu}$ are infinitesimal.

The perturbations can depend on the $z-$coordinate and the $z-$components
of the vector fields can also be non-zero.
{}From (1)
the relevant $z-$dependent terms in the integrand are:
$$
          \half G_{i3} ^a G_{i3} ^a + \half F_{Bi3} F_{Bi3}
          + (D_3 \phi ) ^{\dag} (D_3 \phi )
\eqno (9)
$$
This contribution to the energy
is strictly non-negative and is minimized (that is, made to vanish)
by setting the $z-$components of the gauge fields
to zero and also considering the perturbations to be independent of the
$z-$coordinate. For this reason,
we shall drop all reference to the $z-$coordinate
in the calculations below and it will be understood that the energy is
actually the energy {\it per unit length} of the string.

Now we write (1) after
discarding terms of cubic and higher
order in the infinitesimal perturbations. We find,
$$
\eqalign{
E = E_{NO}[f,v] + \delta E_{NO}[f,v; \phi_2,\delta Z ] & +
E_1 [f,v;\phi_1]+
\cr
&
E_c [f,v;\phi_1,W^{\bar a}] + E_W [f,v;W^{\bar a}, A]
\cr
}
\eqno (10)
$$
where, ${\bar a}=1,2$,
$E_{NO}$ is the energy of the Nielsen-Olesen string and $\delta E_{NO}$ is
the energy variation due to the perturbations $\phi_2$ and $\delta Z^\mu$.
The variation $E_1$ is due to the perturbation $\phi_1$ in the upper
component of the Higgs field:
$$
E_{1} = \int d^2 x \left [
       |{\bar d}_j \phi_1 |^2 + 2 \lambda ( f^2 - \eta^2 /2 ) | \phi_1 |^2
                         \right ] \  ,
\eqno (11)
$$
where,
$
{\bar d} _j \equiv \partial _j - i{\alpha \over 2} cos(2\theta_W ) Z_j \  .
$
The contribution from the $\phi$ and
${\vec W}^{\bar a}$ interaction is:
$$
E_c = cos\theta_W \int d^2 x J_j ^{\bar a} W_j ^{\bar a}
\eqno (12)
$$
$$
J_j ^{\bar a} \equiv \half i \alpha \left [
                     \phi^{\dag} \tau^{\bar a} {\bf d}_j \phi
         - ( {\bf d}_j \phi ) ^{\dag} \tau^{\bar a} \phi \right ]
\eqno (13)
$$
and the energy in the ${\vec W}^{\bar a}$ and $\vec A$ bosons is\refto{error}
$$
\eqalign {
E_W \equiv
&
           \int d^2 x \biggl [
            \gamma {\vec W} ^1 \times {\vec W} ^2 \cdot
                                                 \vec \nabla \times \vec Z
+ \half |\vec \nabla \times {\vec W} ^1
                  + \gamma {\vec W} ^2 \times \vec Z | ^2
\cr
&+
\half |\vec \nabla \times {\vec W} ^2
                  + \gamma \vec Z \times {\vec W} ^1 | ^2
+ \fourth g^2 f^2 ( {\vec W } ^{\bar a} ) ^2
+ \half ( \vec \nabla \times \vec A )^2
                     \biggr ] \  .
\cr}
\eqno (14)
$$
where, $\gamma \equiv g cos\theta_W$. It may be noted that
the $f$ and $\vec Z$ fields in eqs. (11)-(14)
are the unperturbed fields of the string since
we are only keeping upto quadratic terms
in the infinitesimal quantities.
Also, note that the current $J_j ^{\bar a}$
is first order in the perturbation
$\phi_1$ because the $\tau^{\bar a}$ matrices are off-diagonal
and mix the upper and lower components of the Higgs doublet.
That is: $(0,1)\tau ^{\bar a} (0, 1)^T = 0$.

The perturbations of the fields that make
up the string do not couple to the other available perturbations,
i.e. the perturbations in the fields $f$ and $v$ only
occur inside the variation $\delta E_{NO}$. However, we know that
the Nielsen-Olesen string with unit winding number is stable to
perturbations for any values of the parameters. Therefore,
necessarily, $\delta E _{NO} \ge 0$ and
the perturbations $\phi_2$ and $\delta Z^\mu$ cannot destabilize
the vortex. Then we are justified in ignoring these perturbations
and setting $\delta E _{NO} = 0$.

Also note that the variation in the energy vanishes to linear
order in the perturbations.
Therefore the vortex solution given in (2) extremizes the
energy and is a solution of the Weinberg-Salam model regardless
of the values of the parameters in the model.

We now consider an expansion of the remaining perturbations in
Fourier modes.
This gives,
$$
\phi _1 = \chi^m (r) e^{im\theta}
\eqno (15)
$$
for the $m^{th}$ mode where $m$ is any integer. For the gauge fields
we have,
$$
{\vec W}^1 = \left [
\left \{ {\bar f}^n_{1} (r) cos(n\theta ) + f^n_{1} sin(n\theta ) \right \}
{\hat e} _r +
{1 \over r} \left \{
           - {\bar h}^n_{1} sin(n\theta ) + h_{1}^n cos(n \theta ) \right \}
{\hat e} _\theta
\right ]
\eqno (16)
$$
$$
{\vec W}^2 = \left [
\left \{ - {\bar f}^n_{2} (r) sin(n\theta ) + f_{2}^n cos(n\theta ) \right \}
{\hat e} _r +
{1 \over r} \left \{
           {\bar h}_{2}^n cos(n\theta ) + h_{2}^n sin(n \theta ) \right \}
{\hat e} _\theta
\right ]
\eqno (17)
$$
for the $n^{th}$ mode where $n$ is a non-negative integer.
The functionals $E_1$, $E_c$ and $E_W$ may now be expressed in terms of the
modes, $\chi^m$, $f_i^n$ and $h_i^n$. Our goal is to focus on $\delta
E=E_1+E_c+E_W$ and obtain the parameter space $(\beta \equiv
8\lambda / \alpha^2 ,\theta_W)$ for which there are no modes with
$\delta E <0$.
Here we will only sketch the basic steps of the calculation. The full
calculation will be presented elsewhere \refto{mjlptv92}.

It may be shown by examining the form of $\delta E$, that we only need to
look at the $m=0$ and $n=1$ modes to consider the stability of the electroweak
string. This is because these are the modes with the most negative
contribution to the energy variation at the center of the string.
Inserting (15)-(17) in $\delta E$, we find
that the stability problem in the barred functions
completely separates from the stability problem in the unbarred variables.
In addition, it may be shown that if the string is stable to perturbations in
the unbarred variables it will also be stable to perturbations in the barred
variables. Therefore, we are left with five perturbations:
$\chi^0$, $f^1_1$, $f^1_2$,
$h^1_1$ and $h^1_2$. (In what follows we will drop the mode (upper) indices for
simplicity.)

We now define
$$
F_{\pm} = {{f_2 \pm f_1} \over {2}}
\eqno (18)
$$
$$
\xi_{\pm} = {{h_2 \pm h_1} \over 2}
\eqno (19)
$$
$$
\zeta = (1-\gamma v) \chi + \half g f \xi_+
\eqno (20)
$$
After a lot of algebra, we find that the energy variation
is
$$
\eqalign{
\delta E = 2\pi \int dr r \biggl [ \biggl \{
  {{\zeta '}^2 \over {P_+}} + & U(r) \zeta ^2 \biggr \} +
\cr
&
{\rm sum \ of \  whole \  squares} \biggr ]
\cr
}
\eqno (21)
$$
where, primes denote differentiation with respect to $r$,
$$
P_+ = (1 - \gamma v )^2 + \half g^2 r^2 f^2 \  ,
\eqno (22)
$$
$$
U(r) = {{{f'}^2} \over {P_+ f^2}} + {{2 S_+} \over {g^2 r^2 f^2}} +
          {1 \over r} {d \over {dr}} \biggl (
                     {{r f'} \over {P_+ f}} \biggr ) \  ,
\eqno (23)
$$
and,
$$
S_+ = {{g^2 f^2} \over 2} - {{\gamma ^2 {v'}^2} \over {P_+}} +
      \gamma r {d \over {dr}} \left [
                  {{v'} \over r} {{(1-\gamma v)} \over {P_+}}
                             \right ] \   .
\eqno (24)
$$
The sum of whole squares in (21) can be made to vanish by suitably
choosing $F_\pm$ and by setting $\xi_- = 0$. This simply leaves us
with a problem in $\zeta$.

The functional $\delta E$ is in a form
ready to be treated as an eigenvalue problem. That is,
upon performing an integration by parts, we can write
$$
\delta E[ \zeta ] = 2\pi \int dr \  r
              \zeta {\bf O} \zeta
\eqno (25)
$$
where, {\bf O} is the differential operator:
$$
{\bf O} = - {1 \over r} {d \over {dr}} \left (
             {r \over {P_+}} {d \over {dr}} \right ) + U(r)
\eqno (26)
$$
The question of stability now reduces to asking if the operator
{\bf O} has negative eigenvalues in its spectrum.
Therefore we have to determine if the eigenvalue $\omega$ of the
Schrodinger equation,
$$
{\bf O} \zeta = \omega \zeta \  ,
\eqno (27)
$$
can be negative. The eigenfuntion $\zeta$ must also satisfy the
boundary conditions
$\zeta (r=0) = 1$
and $\zeta \rightarrow c$ ($c$ is some constant) as $r \rightarrow
\infty$.

In this way we have reduced the stability problem to a single
eigenvalue problem given by the differential equation in (27)
and the corresponding boundary condition. This problem can be
solved numerically. But before putting the problem on the computer,
we rescale the variables
and the coordinates so that the problem only has two free parameters:
$\beta$ and $\theta _W$. These rescalings are standard in the
literature and may be found in Ref. \cite{tv92}.

The eigenvalue problem in eq. (27) was solved by using a fifth order
Runge-Kutta algorithm.
We kept $\beta$ fixed and found $\theta_w$ for which the
lowest eigenvalue changes sign. We repeated this procedure for several
values of $\beta$ and found the corresponding values of critical parameters
$(\sqrt{\beta}, \sin^2 \theta_w)$.
The above method was used to scan the range $0.07 \leq \beta \leq 1.0$.
Lower values of $\beta$ make the numerical analysis
fairly intensive since then there are two widely different scales in the
problem corresponding to the two widely different masses.
Our results are shown in Fig. 1  where we
plot the critical values of $\sqrt{\beta}$ (the ratio of the Higgs mass
to the Z
mass) versus the corresponding values of $\sin^2 \theta_w$.
In sector III, on
the right-hand side of the data line,
equation (27) had no negative eigenvalues implying
string stability.
Thus we may distinguish three sectors in Fig. 1:
sector I where the electroweak strings are unstable, sector III
where strings are stable, and,
the presently unexplored region shown as sector II
($\beta < 0.07$ or $m_H < 24 GeV$).
It is evident that the
physically realized values: $\sin^2 \theta_w=0.23$ and
$\sqrt{\beta}=m_H / m_Z > 0.62$
(see Ref. \cite{js}) lie entirely inside sector I.
This brings us to the main result of this paper: if the standard electroweak
model is the physically realized model, then the existing
vortex solutions in the bare model are unstable.

Before closing, we would like to point out that even if the
vortex solutions are
unstable, their presence may still be felt in various scattering experiments
since closed
loops and finite string segments could show up as intermediate states.
However, it is more exciting to consider the possibility that
Nature may have chosen
an extension of the standard electroweak model
in which stable vortices are present.

\noindent {\it{Acknowledgements:}}
We would like to thank Miguel Ortiz for his help with the numerical
calculations. MJ acknowledges a SERC Research Studentship.
LP was supported by a CfA postdoctoral fellowship, TV
was supported in part by the National Science Foundation under Grant
No. PHY89-04035 and would like to thank the ITP, Santa Barbara where
part of this work was done.

\references

\item{{*}} Visiting Scientist, Department of Physics, Brown University,
Providence, RI 02903.

\refis{tv91} T. Vachaspati, Phys. Rev. Lett. {\bf 68}, 1977 (1992).

\refis{tv92} T. Vachaspati, TUTP-92-3.

\refis{swas} S. Weinberg, Phys. Rev. Lett. {\bf 19}, 1264 (1967);
A. Salam in ``Elementary Particle Theory'', ed. N. Svarthholm, Stockholm:
Almqvist, Forlag AB, pg 367.

\refis{tvaa} T. Vachaspati and A. Ach\'ucarro, Phys. Rev. D {\bf 44},
3067 (1991).

\refis{jct} J. C. Taylor, ``Gauge Theories of Weak Interactions'',
Cambridge University Press, 1976.

\refis{hnpo} H. B. Nielsen and P. Olesen, Nucl. Phys. B{\bf{61}}, 45 (1973).

\refis{js} J. Steinberger, Phys. Rep. {\bf 203}, 345 (1991).

\refis{mh} M. Hindmarsh, Phys. Rev. Lett. {\bf 68}, 1263 (1992);
G. W. Gibbons, M. Ortiz, F. Ruiz-Ruiz and T. Samols,
DAMTP preprint; A. Ach\'ucarro, K. Kuijken, L. Perivolaropoulos and
T. Vachaspati, CfA 3384 (1992).

\refis{com1} Y. Nambu, Nucl. Phys. B{\bf 130}, 505 (1977);
N. S. Manton, Phys. Rev. D{\bf 28}, 2019 (1983);
M. B. Einhorn and R. Savit, Phys. Lett. B{\bf 77},
295 (1978); V. Soni, Phys. Lett. B{\bf 93}, 101 (1980); K. Huang
and R. Tipton, Phys. Rev. D{\bf 23}, 3050 (1981);
J. M. Gipson and C-H. Tze, Nucl. Phy. B{\bf 183}, 524 (1981).

\refis{gscr} ``Solitons and Particles'', ed. C. Rebbi and G. Soliani,
World Scientific, 1984.

\refis{mjlptv92} M. James, L. Perivolaropoulos, T. Vachaspati,
in preparation.

\refis{mistake} In an earlier version of this paper, we found several
errors that we have now corrected. The resulting map of parameter space,
however, does not change significantly.

\refis{error} In an earlier version of this paper, a factor of 2 had been
omitted in terms 1, 2, 3 and 5 of eq. (14).

\endreferences

\vfill
\eject

\beginsection{Figure Caption}

A map of parameter space showing the region of stability (III) of the
vortex solution. The solutions in Sector I are unstable and we have
not explored the solutions in Sector II
($\sqrt{\beta} < 0.26$).

\vfill
\eject

\endjnl
\end